# Energy Efficient Knapsack Optimization Using Probabilistic Memristor Crossbars

Jinzhan Li[1], Suhas Kumar[2]*, Su-in Yi[1]*

[1]*Dept. of Electrical and Computer Engineering, Texas A&M University, College Station, Texas, USA*

[2]*Rain AI, San Francisco, California, USA*

**Corresponding author. Email: su1@alumni.stanford.edu; yisuin@tamu.edu*

*Constrained optimization underlies crucial societal problems, for instance, stock trading and bandwidth allocation. However, it is often computationally hard, in that complexity grows exponentially with problem size. The big-data era urgently demands low-latency and low-energy optimization at the edge, which cannot be handled by digital processors due to their non-parallel von Neumann architecture. Recent efforts using massively parallel hardware (e.g., memristor crossbars and quantum processors) employing annealing algorithms, while promising, have handled relatively easy and stable problems with sparse or binary representations, such as the max-cut or traveling salesman problems. However, most real-world applications embody three features, which are encoded in the knapsack problem, and cannot be handled by annealing algorithms – dense and non-binary representations, with destabilizing self-feedback. Here we demonstrate a post-digital-hardware-friendly randomized competitive Ising-inspired (RaCI) algorithm performing knapsack optimization, experimentally implemented on a foundry-manufactured CMOS-integrated probabilistic analog memristor crossbar. Our solution outperforms digital and quantum approaches by over 4 orders of magnitude in energy efficiency.*

## Introduction

Optimization problems are crucially important in many applications, such as stock trading and bandwidth optimization (*1-4*). In both these examples, it is necessary to obtain an optimal outcome in near real-time, under several constraints. Nearly all such problems are non-deterministic polynomial-time (NP)-hard, meaning that their complexity grows exponentially with the problem size. As such, solving these problems within a reasonable time, even at small scales, is exponentially resource hungry. Prevailing digital processors, though the default workhorse in addressing such problems, are extremely inefficient. This inefficiency arises from their inherently limited parallelism, owing to the separation of memory and computing within the widely used von Neumann architecture (*5, 6*). Such limited parallelism means that it takes much longer to explore a solution space that depends exponentially on the problem size, thereby needing data-center-scale resources even for small problems. The era of big data accompanies a crucial need to process such data at their source, namely, in edge devices such as cell phones or control computers.

Optimization problems are typically represented by generic formulations with varying levels of complexity (*1*). For instance, the max-cut problem, where the goal is to identify a partitioning of a graph to maximize the number of edges between the two partitions, has no constraints. The max-cut is often represented by a symmetric Hamiltonian matrix using binary weights with arbitrary density of weights, with zero diagonals (*7*). As another example, the traveling salesman problem aims to identify the shortest path that a salesman needs to take to visits all the given cities, under several constraints. For example, two cities cannot be visited at the same time, and the same city should not be visited twice. In contrast to the max-cut problem, the traveling salesman problem encodes constraints and non-binary weights, and its Hamiltonian is extremely sparse (*8*). However, similar to the max-cut problem, its Hamiltonian is symmetric with a zero diagonal. In zero-diagonal systems, the feedback of a specific unit to itself is zero, which makes them inherently stable to feedback and monotonically descending in Hopfield energy with neuronal dynamics (*9, 10*). As such, they are compatible with annealing or cost-minimization algorithms (*11*), which employ feedback-driven updating of the units representing the Hamiltonian.

To address the crucial need to accelerate optimization, the past decade has seen efforts to combine post-digital processors with annealing algorithms (*7, 12-14*). Post-digital processors employ increased parallelism. For instance, memristor crossbars parallelize vector-matrix multiplications (VMMs) to a single clock cycle, which is a notable speed-up compared to semi-parallel digital processors (*15-18*). Similarly, quantum processors employ parallelism within the underlying quantum physics (*14, 19*). Such post-digital hardware is compatible with annealing schemes, such as Hopfield energy minimization, which heavily rely on VMMs. As such, there have been various efforts to solve max-cut (*7, 9*), traveling salesman (*20, 21*), and satisfiability (SAT) (*22*) problems by employing annealing algorithms implemented on memristor crossbars and quantum annealers, especially the D-wave system (*14*). In annealing schemes, a "cost" or an "energy" associated with the problem is minimized, where the cost is defined (based on the Hamiltonian) such that the minimal cost corresponds to the most optimal solution (*8*). Then, by updating the states of the input units, or "neurons", of the system, via both VMM with the Hamiltonian and a feedback mechanism, the cost is minimized. Since the cost function is often non-convex, containing many local minima, reaching the lowest-cost minimum is a probabilistic process.

A key limitation of prevailing approaches is that most real-world problems do not conform to the simplistic representations addressed thus far (*7, 21, 22*). Most problems embody dense representations, non-binary weights that may require high precisions, asymmetry, and, importantly destabilizing self-feedback with non-zero diagonals in the Hamiltonian. The knapsack problem embodies all of these features, and is not addressed by prior methods. The knapsack problem aims to fill a size-limited knapsack, or a container, with objects of varying value and sizes, with the objective of maximizing the value within the container, with the constraint of not exceeding its size limit (*8*) (Fig. 1A). As examples of the knapsack problem, the stock portfolio optimization problem and the bandwidth optimization problem both aim picking objects, such as stocks or channels, with varying sizes, represented by cost or required bandwidth, to maximize value , represented by profits generated by the stocks or channels, with the constraint of not exceeding a size limit, represented by total budget or available bandwidth (*2, 3*). Both examples, stock trading and bandwidth optimization, require near-real-time solutions, which demand edge processing solutions. Annealing schemes are fundamentally incompatible with the destabilizing nature of these problems. Specifically, Hopfield network dynamics with constraints on the non-zero diagonal can disrupt the ground truth (*9*). In addition, the annealing schemes are practically incompatible with the increased complexity posed by their dense and high-precision representations. Hence, we moved towards direct calculations of energy instead of the Hopfield network dynamics (Fig. 1B-C).

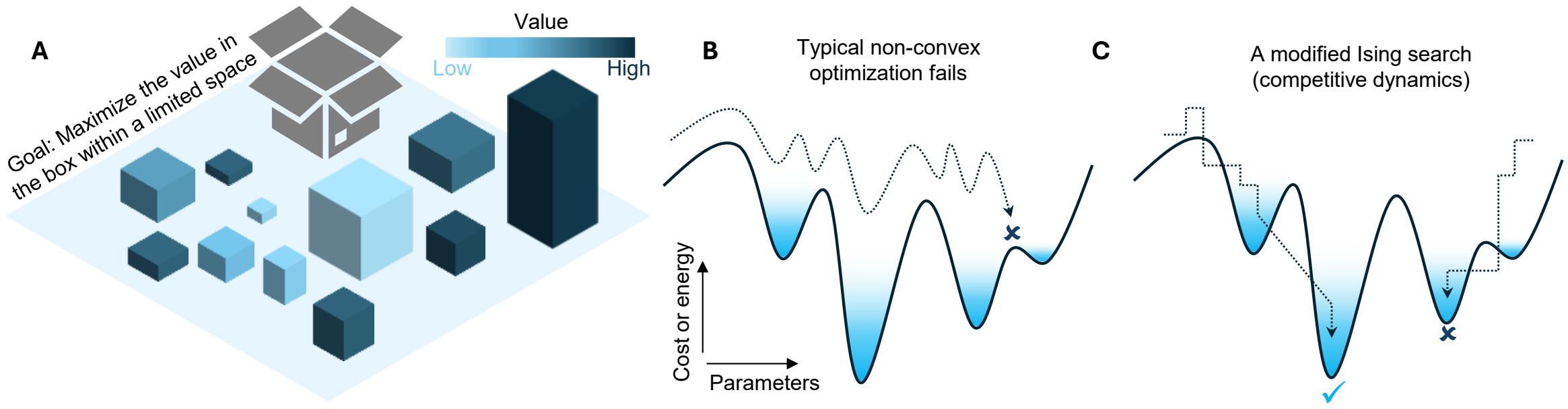

**Fig. 1. Overview of the problem and our approach.** (**A**) Illustration of the Knapsack Problem. (**B**) Neuronal dynamics during typical annealing. (**C**) Neuronal dynamics in our Ising-inspired algorithm.

Here, we present the RaCI algorithm - an Ising-formulation-inspired algorithm, which employs randomness, similar to basin hopping, and competitive dynamics to achieve two outcomes – navigation of a highly non-convex and large search space, and more importantly, avoiding reliance on feedback-based annealing schemes. The randomized competitive Ising conducts searches through a tunneling process that is based on a maintained random flipping schedule until the global minimum. We show that the random feature of the algorithm is not only friendly to be implemented on probabilistic memristor crossbars with finite noise, but can also exploit the noise to guide the randomness. By experimentally implementing the algorithm on an end-to-end foundry-manufactured CMOS-integrated analog memristor hardware with 64×64 memristor crossbars, we demonstrate an improvement in energy efficiency by two-to-four orders of magnitude relative to best central processing units (CPUs), graphics processing units (GPUs), and quantum annealers, all running their most favorable algorithms.

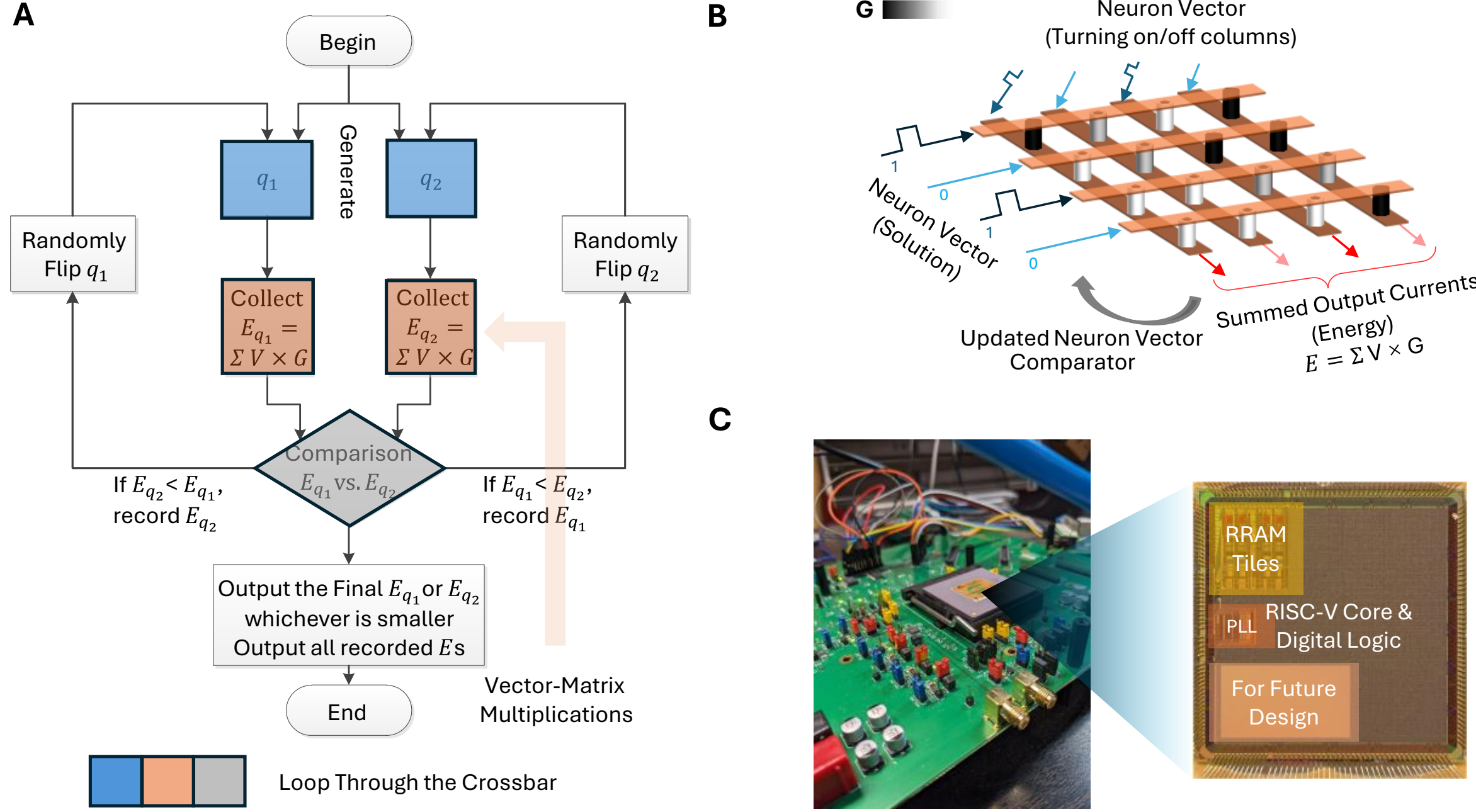

**Fig. 2. Experimental platform and the algorithm-hardware workflow.** (**A**) Experimental Algorithm Workflow. Two initial random neuron vectors $q_1$ and $q_2$ are generated at the beginning, then applied to the crossbar, and looped through the workflow frame until the minimum energy level is reached. (**B**) Schematics of the RRAM

crossbar with steps performed. The neuron vectors apply to the crossbar as voltages and the summed currents are output as energy. (**C**) Photographs of the hardware SoC and experimental platform.

**The randomized competitive Ising-inspired algorithm (The RaCI Algorithm)**

Given *N* items with index $i$, carrying a value of $v_i$ and the corresponding size $w_i$, the objective of the knapsack problem is to maximize the total value filled in the container within its volume limit *W*. Suppose $x_i$ represents the inclusion ($x_i = 1$) or exclusion ($x_i = 0$) of the $i^{\text{th}}$ item, and $y_j = 1$ if the final volume of the container is $j$ ($y_j = 0$ otherwise), for $1 < j < W$. This formulation can be expressed as an energy *E* (Equation (1)), the minimum of which represents the solution to the problem ($\sigma_1$ to $\sigma_3$ are constants) (*8, 23*).

$$E = -\sigma_1 \sum_{i=1}^{N} v_i x_i + \sigma_2 \left(1 - \sum_{j=1}^{W} y_j\right)^2 + \sigma_3 \left(\sum_{j=1}^{W} j y_j - \sum_{i=1}^{N} w_i x_i\right)^2 \qquad \text{Equation}(1)$$

*E* is a function of a binary neuron vector $q = [x_1\, x_2 \ldots x_N\, y_1\, y_2 \ldots y_W]$, where $E = q\, H\, q^{\mathrm{T}}$. $H$ is the Hamiltonian weight matrix derived from the definitions of *E* and $q$ (derived in Supplementary Information Sec. 1). A solution to the problem is found by identifying $q$ corresponding to the minimum *E*. The first negative-coefficient term rewards an increase in total value by reducing *E* and minimizing the second positive-coefficient term ensures a specific total size not exceeding *W*. Minimizing the third positive-coefficient term ensures that the total size represented by $y_j$ matches the total size represented by $w_i x_i$. Thus, by minimizing *E*, we can identify all elements of $q$. Length of the vector $y$ can be reduced from *W* to $1 + W/s$ (*s* denotes a shrink factor) or $1 + \log(W)$ by using encoding schemes (*8*). As such, in our implementation, we used 15 total neurons, corresponding to 5 objects (*N* = 5) and the size limit 10 (*W* = 10).

The RaCI algorithm first calculates the weight matrix, or the Hamiltonian *H*, from the given problem definition. Next, two sets of competitive neuron vectors ($q_1$ and $q_2$) are generated (Fig. 2A), where only the very initial $q_2$ is a randomly flipped version of $q_1$. Then, the algorithm conducts a search for an optimized solution through random flipping of the neuron bits within$q_1$ or $q_2$. A random number of neurons, *n*, to be flipped is decided, following which *n* neurons are randomly selected to be flipped. After each flipping, the new energies $E_1$ and $E_2$ are calculated, corresponding to $q_1$ and $q_2$. The lower of the energies is selected, and stored as the current minimum energy, along with its neuron state. This process continues until the energy saturates at a minimum or a pre-determined number of iterations is reached.

The RaCI algorithm borrows from multiple previous optimization efforts (*23, 24*). There are similarities to basin hopping, where neurons are flipped at random, often during annealing. The use of two competitive neuron vectors, instead of one, enables a balance and prevents the system from getting stuck in unfavorable conditions. That is, this scheme employs a dynamical process in place of traditionally used feedback. The algorithm randomizes both the number of neurons within the neuron vector to be flipped, as well as the indices of the neurons, which together enhance the randomness. In addition, the two neuron vectors are used for comparison purposes where we record the one with a lower energy state and flip the one with a higher energy state. This is considerably efficient when using a single piece of hardware. Doubling the hardware resources will achieve the same outcome, however, it also brings a higher energy consumption as the problem has to be stored in two sets of duplicated hardware. The RaCI algorithm has a similarity to differential evolution in that it evaluates the neuron based on certain conditions, which is the energy in our case (*25*). However, it is different from differential evolution in that RaCI is based on random flipping search for replacing the best solutions, with the help of device noise for randomness. Meanwhile, the general differential evolution will require some population-based crossover for the neuron candidates.

Particularly, the algorithm avoids feedback from the output to the input neurons, since the knapsack problem is fundamentally incompatible with feedback mechanisms (*11*). Further, as we will show later, the natural analog hardware noise from the memristor crossbars can be used to naturally implement the randomization in the algorithm. The noisy hardware nature also makes it rare for two suboptimal neurons to stay at the same energy experimentally due to the random noise in the array, which helps maintain momentum of the search and flipping process.

**Hardware implementation**

For hardware implementation, we used the Cerebro-2 chip, which contains four 64×64 memristor crossbar arrays fully integrated with CMOS control circuits, and end-to-end foundry manufactured at the 65 nm technology node (*26*). The RRAM stack was manufactured with a wide-band dielectric layer and a metal oxide layer using physical vapor deposition and atomic layer deposition (*26*). The diameter of the bitcells is 200 nm (*27*). The devices are capable of analog tunability (*28*), with up to 7 bits of precision, and finite circuit noise typical of most memristor hardware. As such, the hardware was compatible with the RaCI algorithm (FIG. 2B-D). To implement the algorithm, input vector $q$ is presented as voltages to the rows, while matrix $H$ is programmed as the conductance of the crossbar array. $q^{\mathrm{T}}$ is presented as a gate voltage to the columns, wherein the gate voltage turns an entire column on or off, such that no current flows in the off state. Thus, the output vector obtained by Ohms law and Kirchoff's laws, following summation of the individual elements, represents the scalar energy $E = q\,H\,q^{\mathrm{T}}$. Thus, our highly functional hardware, with simultaneous access to the column gating and VMMs, enables direct calculations of $E$ at every clock cycle, without the need for any external processing. This solution is unlike prior efforts, which could perform only VMMs in memristor crossbar arrays and needed external circuits to calculate $E$.

Since our weight matrices are asymmetric, we program half of an array with the positive weights and half of another array with negative weights (Fig. 3A). Remaining arithmetic and storage functions are supported in the on-chip peripheral circuits or in off-chip digital circuits. For example, in the case of summing up the specific currents, corresponding to Hamiltonian energies, collected from the crossbar, it was performed through peripherals (Raspberry Pi) in the digital domain. The results from the positive and negative arrays are subtracted to obtain the final result. Notably, since we calculate the energy directly from the programmed matrix, and not from an ideal matrix as is typical in Ising and Hopfield algorithms, the resulting energy is bound to contain analog circuit noise affecting both the readouts and the weights, even when the neuron input does not change (Fig. 3B). As such, for a given solution (neuron state), the resulting energy lies in a band defined by hardware fluctuations (Fig. 3C). The read noise is a known behavior of most filamentary-switching RRAMs. Specifically, defects and traps near the filament can capture and release electrons, changing the local conduction pathways. Noise may be mitigated to a limited extent using bulk-switching RRAMs (*27*). However, in our application, we are able to exploit the noise to promote randomness. To perform simulations, we modeled the devices to exhibit state-independent uniform noise, which is justified by prior measurements of nominally identical devices (*29, 30*).

With the above procedures and conditions, the dynamically evolving solution to a 15-neuron (225 weights or synapses) knapsack problem (Fig. 3D) exhibits the known signature of energy minimization, similar to annealing. The energy saturates at a minimum band with finite fluctuations within 40 iterations, which is the number of times the neuron vector is updated. We repeated the experiments 100 times for each variation in the number of iterations. The probability $p$ of finding the globally optimal solution increased monotonically with the number of iterations (Fig. 3E), as expected. However, it is also apparent that at larger numbers of iterations, the impact on $p$ diminishes. To quantify this value, we calculated the number of times the experiment needs to be repeated, $r$, to obtain $p$ = 0.99 (which is calculated via the probability of failure by repeating the process $r$ times: $(1-p)^r = 1 - 0.99$ (*9*)). We then calculate the total resource expended by calculating the total iterations multiplied by $r$. The trend in this quantity (Fig. 3F) exhibits an interesting behavior. It indicates that The resource spent in solving the problem is minimal when performing fewer iterations repeatedly or a large number of iterations repeated very few times. The former case is possibly friendlier to edge applications, where latency is important and many minor variations of the problem may have to be solved repeatedly.

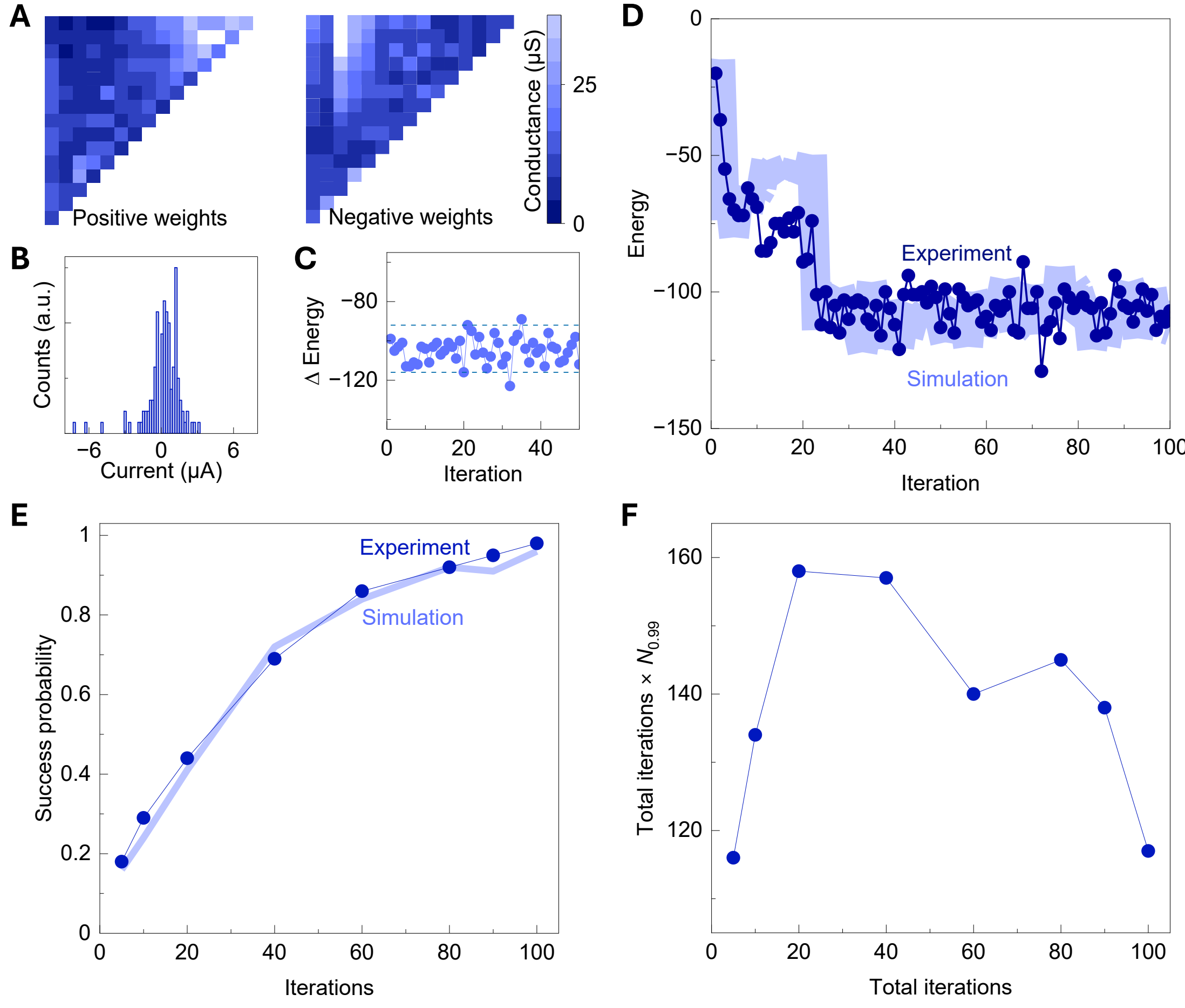

**Fig. 3. Experiemental implementation of RaCI on RRAM crossbars.** (**A**) Experimental read-out matrices after programming. The positive and negative matrix represent the positive and negative weights, respectively. (**B**) The histogram of errors. (**C**) Neuron noise: the energy fluctuations of hardware for the same neuron during 50 times energy reading. (**D**) Neuron energy minimization. The experimental neuron energy trajectory with the increased number of iterations for one random successful case that reaches the optimal solution. The experimental energy and the corresponding simulated energy of the neurons recorded in the hardware experiement are on top of each other. The energy level of the best solution corresponds to a fluctuated energy window due to the RRAM crossbar noise. (**E**) Experiemental success probability vs. iterations. The experimental and simulation success rate of solving the knapsack problem over different discrete iteration numbers ranging from 5 to 100. (**F**) The total iteration number for >99% success probability of solving the problem based on experimental success probablity. The total iteration number represents the minimum best iteration number for solving the problem with 99% successful rate. The combination of iteration number and success probability ($p$) determines the repetition number with different intial neuron vectors to reach the optimal problem solution with 99% probablity based on $N_{0.99} = \log_{(1-p)} 0.01$. The total iteration number is based on the multiplication of the iteration number and $N_{0.99}$.

## Performance under noise and scalability

Since the RaCI algorithm is designed to be friendly to analog and edge hardware, it is important to understand its resilience to noise. We studied the effects of noise on the algorithm (Fig. 4A), with 'native' noise being identical to the native hardware noise. The results show that noise is generally harmful to the algorithm. However, for native levels of noise, there was no measurable deviation from the theoretical limits (zero noise). A detectable change was observed with 3× the level of native noise.

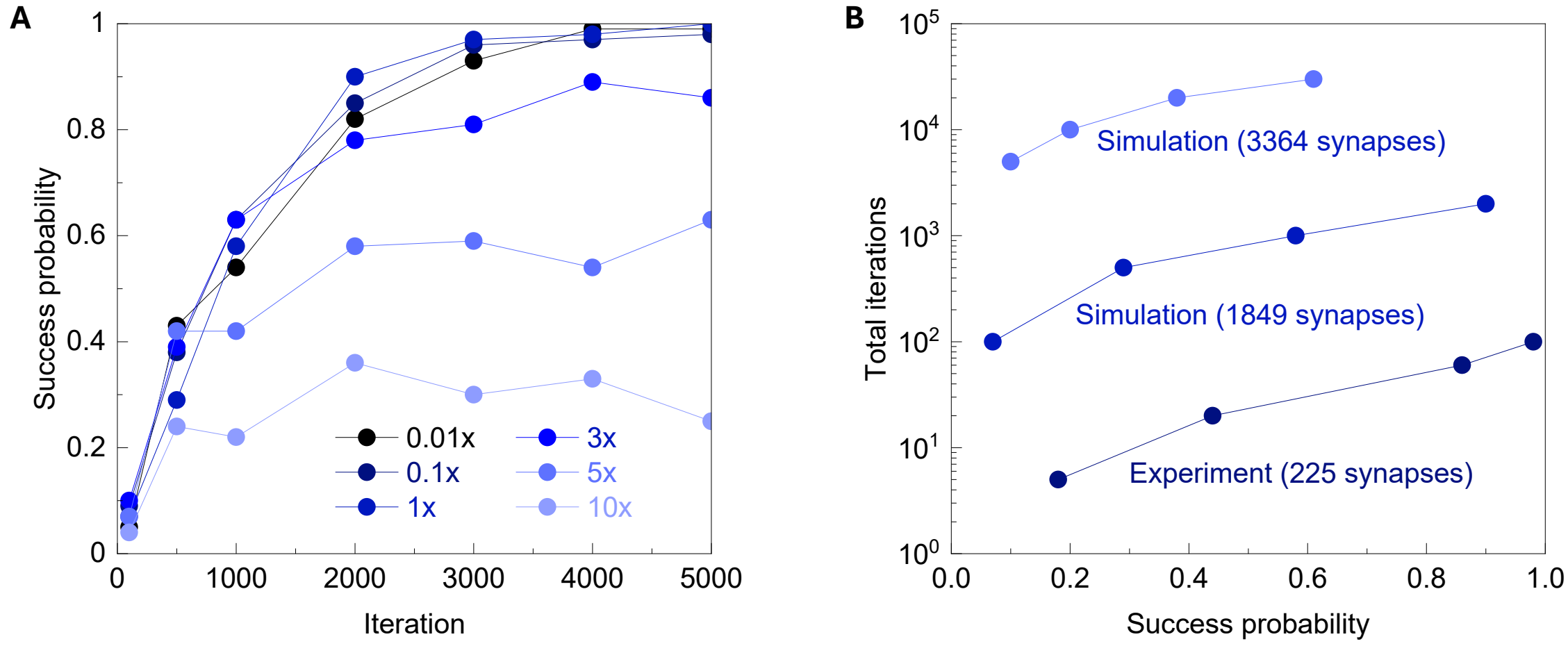

**Fig. 4 Effect of different conditions and scalability of the problem.** (**A**) Effect of different conditions for the knapsack problem under 7-bit precision. The success probability of a scaled Knapsack problem with 1849 synapses over different discrete iteration numbers under the cases where up to 10× noise is applied to the crossbar for simulation. (**B**) Success probability for scaling beyond sizes with 1849 synapses and 3364 synapses.

To study such scalability, we calculated the number of iterations required to achieve different success probabilities in problems with 43 neurons (1849 weights, 10 objects), which exhibited an increase in the required number of iterations by roughly one order of magnitude, while employing a native magnitude of noise. This result is unsurprising. However, when we attempted to solve a problem with 58 neurons, we found a near-zero success probability for any number of iterations up to 30,000. We hypothesized that as the problem size grew, the energy landscape will be more complex, leading to a more susceptible system, which may be more prone to noise. A problem with 58 neurons (3364 weights, 15 objects) likely exceeded that limit. In such cases, we will need noise mitigation schemes. As one such scheme, we turned to a recently discussed method (*31*), wherein multiple devices encoding the same information are measured and averaged, thereby minimizing noise (*18, 31*). We used two additional crossbars as references, in addition to the original crossbar, each programmed with a nominally identical matrix. Thus, by using three devices to represent every weight, and averaging the measured conductance of the three devices, we were able to reduce the noise to a notable extent. Using this scheme, the 58-neuron problem was more tractable. With up to 30,000 iterations, we were able to obtain a success probability of 0.65.

## Using hardware noise to drive the randomness in the RaCI algorithm

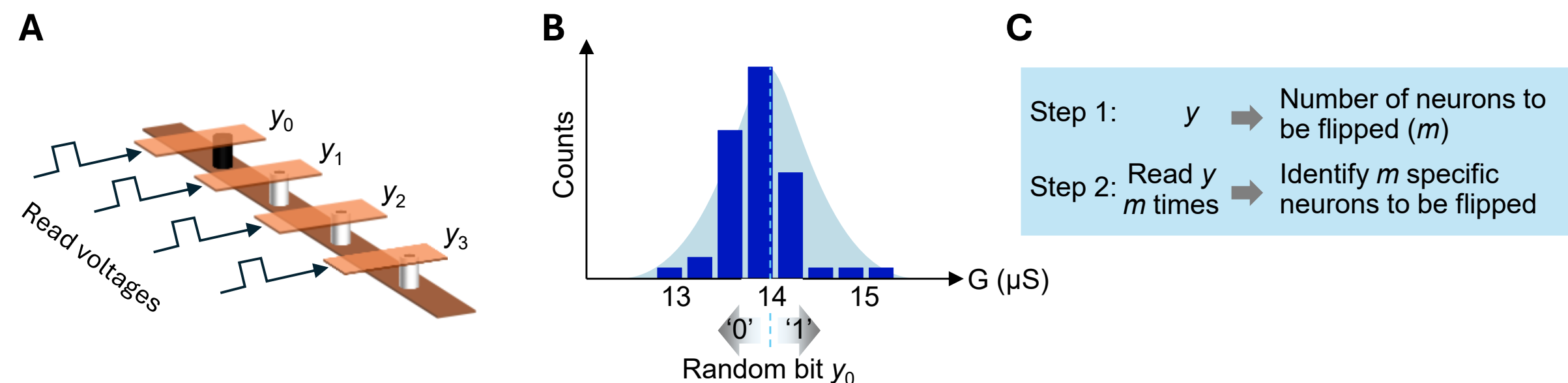

**Fig. 5. Randomness implementation of RaCI algorithm with hardware noise.** (**A**) Hardware-based RaCI decision maker: A scheme for generating the flipping sequence based on hardware noise. A number of devices with natural hardware noise are read, which can produce a random bit vector $y$, with each element of $y$ corresponding to every device. (**B**) The resulting Gaussian-like profile of the conductance values (experimental data displayed, along with an illustrated Gaussian profile) allows the generation of a random bit, by using the mean of the distribution as a threshold. (**C**) The process of generating $y$ is used to identify both the number of neurons to be flipped and the specific neurons that need to be flipped.

To implement the inherent randomness in the RaCI algorithm, it is possible to use the hardware's native noise. In other words, the analog devices in the memristor crossbar are be used as a random number generator to determine the flipping of the bits, which is an integral part of the algorithm. As an example (Fig. 5), four devices represent the experimental Knapsack problem. Assuming the device noise follows Gaussian distributions, if the readout value of one device is over the mean, then the readout is recorded as "1", otherwise "0". Thus, four devices are used to obtain a four-bit pattern. A random number $m$ ranging from 0 to 5 is generated based on the recorded pattern, indicating the number of bits to be flipped. The process is then repeated $m$ times to determine which $m$ bits in the neuron vector need to be flipped. In case we reach a redundant pattern that has no corresponding number, which only happens with an extremely small probability, we run the process again. In Supplementary Information Sec. 5, we provide additional information on the minimum number of devices needed for this process as the number of neurons increases, and the probability of obtaining redundant information.

## Performance benchmarking and discussions

We performed a performance comparison between our system and a traditional CPU, a GPU, and the D-wave 2000Q quantum system. For RRAM, the performance metrics are obtained from our chip's design scaled to a 16 nm technology node, using leading foundry scaling rules (*7*). Our estimates include all the necessary peripheral components. For a CPU, we used a 20 W power design, and for a GPU, we used a 150 W power design (*7*). Specifically, the CPU data is based on Ref. (*32*) running on an Intel i7 processor with identical size of the problems using the RaCI algorithm. It is worth noting that the RaCI seems to be extremely inefficient on CPU, and it generates the longest annealing time compared to other algorithms (*32*), such as branch and bound (BB), generic algorithm (GA) and dynamic programming (DP). Therefore, we also made a comparison of a CPU running DP. The GPU data are tested on a 1.15 GHz Nvidia Titan X with an identically sized problem running the RaCI algorithm (from Table V in Ref. (*33*)). The performance is adjusted by appropriate scaling factors to represent performance on match the state-of-the-art processors (*34, 35*) (Intel I9-13900K and Nvidia RTX 4080). We also compared with a more efficient branch and bound (BB) approach for GPU implementation (*36*) (data from Ref. (*35*) was adjusted to account for the difference in the problem size). In addition, we compare with D-wave's 2000Q quantum annealer that consists of 2048 qubits (*14*). The data (Fig. 6 and Supplementary Table S1) show that the memristor hardware running the RaCI algorithm on a problem with 15 neurons offers more than four orders of magnitude better energy efficiency compared to the state-of-the-art CPU and GPU systems, and also a D-Wave2000Q quantum approach. The datasets for benchmarking on CPUs and GPUs, from the relevant references cited, were randomly generated with specific parameters, such as volume limits and the number of items. Similarly, we generated randomly generated datasets with similar parameters to benchmark out hardware-algorithm combination. CPU and GPU performances naturally include device-host communication. However, in our case, the entire problem is stored in the RRAM array owing to the in-memory computing scheme. As such, overheads of communicating with a host are not included in our calculations. While we profile our solution for edge use cases, when handling cloud-scale problems, an RRAM hardware too may have to account for device-host communications, which is deferred to another dedicated study.

As we noted in the previous section, larger problems face a challenge when solved using the RaCI algorithm running on memristor crossbars. To quantify and accurately capture such performance issues at larger scales, we benchmarked the energy consumption of solving a problem with 58 neurons using all the approaches we used to benchmark the problem with 15 neurons. It is apparent (Fig. 6) that the resources required to solve the larger problem is similar to those required to solve the 15-neuron problem in the case of CPU and GPU within a factor of 3, while for the quantum annealer, it was within a factor of 100. However, for the memristor crossbar running the RaCI algorithm, the difference was a factor of 10,000. In other words, as the problem size grows, our solution's advantages diminish.

It is possible that at even larger scales (*37*), the energy needed for the RaCI approach on memristor crossbars may be more than the energy needed to perform optimization on CPUs and GPUs. It is also possible that at large scales, the RaCI approach maintains some advantages, though small. Irrespective of those scaling trends (which are complicated to calculate), it is clear that at small scales friendly to edge environments, the RaCI approach running on memristor crossbars is highly beneficial. Notably, our advantages, demonstrated up to 58 neurons (15 knapsack objects), are relevant to edge applications. Increasing the problem size from 15 knapsack objects to 20 objects increases the search space from 30,000 possible solutions to over 1,000,000 possible solutions. As such, the edge applications lie in the vicinity of the size of the problems studied here. Further, it is impractical to deploy GPUs or high-end CPUs on edge devices such as wearable watches, needless to say quantum systems on any edge system. As such, though our solution has a limited scope of utility, it offers a strong and much needed solution in a space that is not served by any other technology.

Regarding possible designs of a chip running the RaCI algorithm, all the multiplications as well as additions can be performed in the analog domain, without conversion to digital information, although our proof-of-principle demonstration employed off-chip peripheral circuits to perform additions, due to ease of implementation. Further, although we included analog-to-digital converters (ADCs) in our chips, to aid demonstrations of multiple algorithms, the RaCI algorithm can be implemented without the use of ADCs. To achieve the comparator function of the Hopfield-like network, a design could include only sample-and-hold circuits or analog comparators following transimpedance amplifiers at the output of the multiplication units. Such circuit components usually consume much lower energy compared to ADCs, making an energy-efficient implementation of the RaCI algorithm plausible.

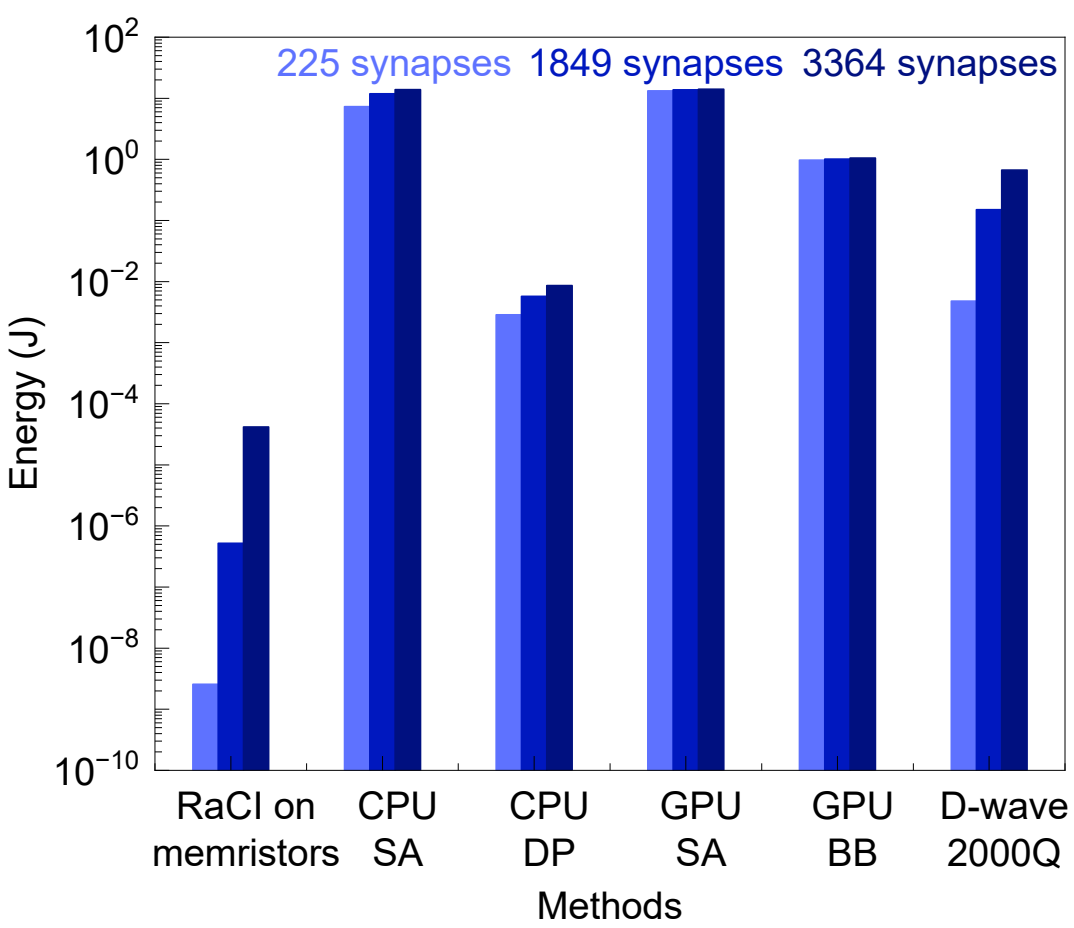

**Fig. 6. Energy benchmark of different systems.** SA represents the RaCI algorithm. DP (dynamic programming) and BB (branch and bound) denote favorable algorithms for CPU/GPU. D-wave 2000Q is a quantum annealer reported in Ref. (*14*). The energy quantities correspond to the energy required to find the optimal solution to the problem, summed over all the necessary iterations.

## Conclusion

In conclusion, we demonstrate that the analog-hardware-friendly RaCI algorithm running on analog memristor crossbars can handle dense and non-binary computationally hard optimization problems with destabilizing connections. Such problems cannot be handled by classical or quantum annealing systems. We showed that our algorithm-hardware solution outperforms CPUs, GPUs and quantum systems by at least two orders of magnitude for edge-scale problems, a space that crucially needs a new solution.

## Acknowledgements

J.L. and S.Y. were supported by the Air Force Office of Scientific Research (AFOSR) under grant no. AFOSR-FA9550-19-0213, titled 'Brain Inspired Networks for Multifunctional Intelligent Systems in Aerial Vehicles'. The authors also thank Applied Materials Inc. for loaning their memristor arrays. Luke Thomas is gratefully acknowledged for support in bringing up the memristor chip.